%% file: main.tex
\documentclass[conference]{IEEEtran}

\IEEEoverridecommandlockouts

\usepackage{cite}
\usepackage{amsmath,amssymb,amsfonts}
\usepackage{algorithmic}
\usepackage{graphicx}
\usepackage{textcomp}
\usepackage{xr}
\usepackage[table]{xcolor}

\def\BibTeX{{\rm B\kern-.05em{\sc i\kern-.025em b}\kern-.08em
		T\kern-.1667em\lower.7ex\hbox{E}\kern-.125emX}}

\usepackage{tikz}
\usetikzlibrary{backgrounds}
\usetikzlibrary{spy}
\usetikzlibrary{calc,arrows.meta}
\usepackage{pgfplots}
\usetikzlibrary{plotmarks}
\usetikzlibrary{arrows.meta}
\usetikzlibrary{positioning,fit}
\usetikzlibrary{intersections}
\usetikzlibrary{patterns}
\usetikzlibrary{decorations.shapes}

\usepgfplotslibrary{patchplots}
\usepgfplotslibrary{colormaps}

\usepackage{cancel}

\usepackage{pdfpages} 

\usepackage{multirow}
\usepackage{makecell}

\usepackage{algorithmic}
\usepackage[ruled,vlined]{algorithm2e}
\SetKwRepeat{Do}{do}{while}

\newcommand{\exportFigures}{true}
\newcommand{\exportFiguresAsPNG}{true}

\ifthenelse{\equal{\exportFigures}{true}}
{
	\usepgfplotslibrary{external}
	\tikzexternalize[prefix=compiled_tikz_figures/,optimize command away=\includepdf]
	\ifthenelse{\equal{\exportFiguresAsPNG}{true}}
	{
		\tikzset
		{   png export/.style={
				external/system call={
					pdflatex \tikzexternalcheckshellescape -halt-on-error --extra-mem-top=10000000 -interaction=batchmode -jobname "\image" "\texsource" && pdftops -eps "\image.pdf" && convert -density 700 -transparent white "\image.pdf" "\image.png"
		}}}
		\tikzset{png export}
	}
	{}
}
{}

\usepackage{url}

\usepackage{bm}
\usepackage[caption=false,font=footnotesize]{subfig}
\usepackage[normalem]{ulem}

\usepackage{mathtools, cuted}

\usepackage{acronym}

\usepackage{booktabs}
		
\usepackage{pdfpages}

\usepackage[latin1]{inputenc}
\usepackage{tikz}
\usetikzlibrary{shapes,arrows}
\usetikzlibrary{arrows.meta}
\usetikzlibrary{positioning}
\usetikzlibrary{spy,backgrounds}
\usetikzlibrary{math}

\usepackage{ellipsis}

\usetikzlibrary{calc}
\usetikzlibrary{decorations.pathreplacing,decorations.markings,shapes.geometric}
\usetikzlibrary{decorations.pathmorphing}
\usetikzlibrary{fit}
\usetikzlibrary{pgfplots.groupplots}

\usetikzlibrary{calc,arrows.meta}

\usetikzlibrary{backgrounds} 

\definecolor{green(pigment)}{rgb}{0.0, 0.65, 0.31}
\definecolor{frenchblue}{rgb}{0.0, 0.45, 0.73} 
\definecolor{mediumcandyapplered}{rgb}{0.89, 0.02, 0.17}

\usepackage[most]{tcolorbox}
\tcbuselibrary{breakable}
\tcbset{every box/.style={enhanced,breakable}}
\tcbset{colframe=black,colback=red!10,enhanced,breakable,sharp corners}

\usepackage{enumitem} 

\usepackage{notation}

\usepackage{adjustbox}

\definecolor{alex}{RGB}{51,183,150}
\definecolor{erik}{RGB}{235,134,52}
\definecolor{Vicky}{RGB}{235,134,52}

\newcommand{\ticked}{$\text{\rlap{$\checkmark$}}\square$}
\newcommand{\unticked}{{$\square$}}
\newcommand{\tick}[1]{\ifthenelse{#1=1}{\ticked}{\unticked}}

\input{./base/definitions.tex}

\newlength{\figureheight}
\newlength{\figurewidth}
\graphicspath{{./figures/}}

\allowdisplaybreaks

\input{./base/plotstyles.tex}

\input{./base/tikzsymbols.tex}

\begin{document}

\title{Multi-Sensor Fusion of Active and Passive Measurements for Extended Object Tracking}

\author{Hong Zhu$^{1,2}$, Alexander Venus$^{1,2}$, Erik Leitinger$^{1,2}$, and  Klaus Witrisal$^{1,2}$
\thanks{The financial support by the Christian Doppler Research Association, the Austrian Federal Ministry for Digital and Economic Affairs and the National Foundation for Research, Technology and Development is gratefully acknowledged.}

\\
\small{{$^1$Graz University of Technology, Austria},
}\\
\small{{$^2$Christian Doppler Laboratory for Location-aware Electronic Systems}}\\
}


\maketitle
\frenchspacing

\renewcommand{\baselinestretch}{0.98}\small\normalsize 

\begin{abstract}
	\input{./sections/abstract.tex}

\end{abstract}

\acresetall 


\IEEEpeerreviewmaketitle



\section{Introduction}\label{sec:introduction}

\input{./sections/introduction.tex}


\section{Radio Signal Model}\label{sec:signal_model}

\input{./sections/signal_model.tex}

 
\section{System Model}\label{sec:system_model}

\input{./sections/system_model.tex}


\section{Ideal Scattering Model}\label{sec:ideal}

\input{./sections/ideal.tex}

\section{Results}\label{sec:results}

\input{./sections/results.tex}


\section{Conclusion and Future Work}\label{sec:conclusion}

\input{./sections/conclusion.tex}


%

\input{./sections/acronyms.tex}





\bibliographystyle{IEEEtran}
\bibliography{IEEEabrv,joint_active_passive_tracking,EOT_and_PDA,signalModelingEA,references}

\end{document}

%% file: base/definitions.tex
\newcommand{\rmv}{\hspace*{-.3mm}}

\newcommand{\vm}[1]{\ensuremath{\bm{#1}}} 

\renewcommand{\Re}[1]{\ensuremath{\text{Re}\!\left[#1\right]}}
\renewcommand{\Im}[1]{\ensuremath{\text{Im}\!\left[#1\right]}}

\newcommand{\norm}[2]{\ensuremath{\lVert #1 \rVert^{#2}}}

\newcommand{\minus}{\rmv - \rmv}

\newcommand{\s}{\hspace*{0.5pt}}

\providecommand{\norm}[1]{\lVert#1\rVert}

\newcommand{\nn}{\nonumber}

\mathchardef\Re="023C
\mathchardef\Im="023D

%% file: base/plotstyles.tex
\definecolor{mycolor01}{rgb}{0.00000,0.00000,1.00000}
\definecolor{mycolor02}{rgb}{0.133,0.545,0.133}
\definecolor{mycolor03}{rgb}{0.50000,0.00000,0.50000}
\definecolor{mycolor05}{rgb}{1.00000,0.83984,0.00000}
\definecolor{mycolor04}{rgb}{0.92969,0.50781,0.92969}
\definecolor{mycolor06}{rgb}{1.00000,0.64453,0.00000}
\definecolor{mycolor07}{rgb}{0.50000,0.50000,0.50000}
\definecolor{mycolor08}{rgb}{1.00000,0.00000,0.00000}
\definecolor{mycolor09}{rgb}{0.2510 ,0.8784, 0.8157}
\definecolor{mycolor10}{rgb}{0.54297,0.00000,0.00000}
\definecolor{mycolor11}{rgb}{0.6445, 0.1641,0.1641}
\definecolor{mycolor12}{rgb}{1, 0, 1}

\pgfdeclarelayer{back1}
\pgfdeclarelayer{lay1}
\pgfdeclarelayer{back2}
\pgfdeclarelayer{lay2}
\pgfsetlayers{back2,lay2,back1,main,lay1}

\makeatletter

\tikzset{
	nomorepostactions/.code={\let\tikz@postactions=\pgfutil@empty},
	decmark/.style 2 args={decoration={markings,
			mark= between positions 0 and 1 step (1/6)*\pgfdecoratedpathlength with{%
				\tikzset{#2,every mark}\tikz@options
				\pgftransformresetnontranslations
				\pgfuseplotmark{#1}%
			},  
		},
		postaction={decorate},
		/pgfplots/legend image post style={
			mark=#1, mark options={#2}, every path/.append style={nomorepostactions}
		},
	},
	markbeginend/.style 2 args={decoration={markings,
			mark= between positions 0 and 1 step (1)*\pgfdecoratedpathlength with{%
				\tikzset{#2,every mark}\tikz@options
				\pgfuseplotmark{#1}%
			},  
		},
		postaction={decorate},
		/pgfplots/legend image post style={
			mark=#1,mark options={#2},every path/.append style={nomorepostactions}
		},
	},
	markend/.style 2 args={decoration={markings,
			mark= at position \pgfdecoratedpathlength with{%
				\tikzset{#2,every mark}\tikz@options
				\pgfuseplotmark{#1}%
			},  
		},
		postaction={decorate},
		/pgfplots/legend image post style={
			mark=#1,mark options={#2},every path/.append style={nomorepostactions}
		},
	},
	posmark/.style 2 args={decoration={markings,
			mark= at position #2 with{%
				\tikzset{solid,every mark}\tikz@options
				\pgftransformresetnontranslations
				\pgfuseplotmark{#1}%
			},  
		},
		postaction={decorate},
		/pgfplots/legend image post style={
			mark=#1,mark options={solid},every path/.append style={nomorepostactions}
		},
	},
}

\makeatother

\pgfplotsset{
	resultStyle1/.style={mark=none, line width=0.5pt, mycolor01, decmark={oplus}{solid}},
	resultStyle2/.style={mark=none, line width=0.5pt, mycolor02, decmark={triangle}{solid}},
resultStyle3/.style={mark=none ,line width=0.5pt, mycolor03, decmark={+}{solid}},
resultStyle4/.style={mark=none, line width=0.5pt, mycolor06, decmark={star}{solid}},
resultStyle5/.style={mark=none, line width=0.5pt, mycolor08, decmark={o}{solid}},
resultStyle6/.style={mark=none, line width=0.5pt, mycolor05, decmark={square}{solid}}, 
resultStyle7/.style={mark=none, line width=0.5pt, mycolor09, decmark={diamond}{solid}}, 
resultStyle8/.style={mark=none, line width=0.5pt, mycolor11, decmark={otimes}{solid}}, 
resultStyle9/.style={mark=none, line width=0.5pt, mycolor12, decmark={x}{solid}}, 
resultStyleBase/.style={mark=none, line width=0.5pt,}, 
compareStyle1/.style={mark=none, line width=0.5pt, mycolor01},
compareStyle2/.style={mark=none, line width=0.5pt, mycolor02},
compareStyle3/.style={mark=none ,line width=0.5pt, mycolor03},
compareStyle4/.style={mark=none, line width=0.5pt, mycolor06},
compareStyle5/.style={mark=none, line width=0.5pt, mycolor08},
compareStyle6/.style={mark=none, line width=0.5pt, mycolor05}, 
compareStyle7/.style={mark=none, line width=0.5pt, mycolor09}, 
compareStyle8/.style={mark=none, line width=0.5pt, mycolor11}, 
compareStyle9/.style={mark=none, line width=0.5pt, mycolor12}, 
}

\pgfplotsset{
compat=newest,
simple style group/.style={
label style={font=\scriptsize},
legend style={font=\scriptsize},
tick label style={font=\scriptsize},
nodes near coords style={font=\scriptsize},
title style={font=\scriptsize},
scale only axis,
grid style={dotted},
mark options={solid}, 
},
simple style/.style={
label style={font=\scriptsize},
legend style={font=\scriptsize},
tick label style={font=\scriptsize},
nodes near coords style={font=\scriptsize},
title style={font=\scriptsize},
width=\figurewidth,
height=\figureheight,
at={(0\figurewidth,0\figureheight)},
scale only axis,
grid style={dotted},
mark options={solid}, 
},
base style/.style={
label style={font=\scriptsize},
legend style={font=\scriptsize},
tick label style={font=\scriptsize},
nodes near coords style={font=\scriptsize},
title style={font=\scriptsize},
width=\figurewidth,
height=\figureheight,
at={(0\figurewidth,0\figureheight)},
scale only axis,
cycle list={
	{mark=none, line width=0.5pt, mycolor01, solid},
	{mark=none, line width=0.5pt, mycolor02, dash dot},
	{mark=none ,line width=0.5pt, mycolor03, densely dashed},
	{mark=none, line width=0.5pt, mycolor04, dash dot dot},
	{mark=x   , line width=0.5pt, mycolor05},
	{mark=.   , line width=0.7pt, mycolor06}, 
	{mark=square,only marks, mark size = 0.8pt, mycolor07,
		mark options = {line width = 0.4pt}},
	{mark=x,     only marks, mark size = 1.3pt, mycolor08,
		mark options = {line width = 0.4pt}},
	{mark=o,     only marks, mark size = 0.8pt, mycolor09,
		mark options = {line width = 0.4pt}},
	{mark=o, mycolor10},
},
grid style={dotted},
xmajorgrids,
ymajorgrids,
mark options={solid}, 
},
base style group/.style={
	label style={font=\scriptsize},
	legend style={font=\scriptsize},
	tick label style={font=\scriptsize},
	nodes near coords style={font=\scriptsize},
	title style={font=\scriptsize},
	scale only axis,
	grid style={dotted},
	xmajorgrids,
	ymajorgrids,
	mark options={solid}, 
},
std graph style new/.style={
xlabel style={yshift=1mm},
ylabel style={yshift=-1.5mm},
yticklabel style={xshift=1mm},
},
color lines style/.style={
cycle list={
	{mark=none, mycolor01, decmark={oplus}{solid} },
	{mark=none, mycolor02, decmark={+}{solid} }, 
	{mark=none, mycolor03, decmark={triangle}{solid} }, 
	{mark=none, mycolor04, decmark={star}{solid} }, 
	{mark=none, mycolor05, decmark={o}{solid} },
	{mark=none, mycolor06, decmark={square}{solid} },
},
},
meas graph style/.style={
xlabel style={yshift=1mm},
ylabel style={yshift=-1mm},
xmajorgrids,
ymajorgrids,
mark repeat = 1,
mark phase = 0,
cycle list={
	{color=black, only marks, mark=*, mark size=0.5pt, mark options={solid, black}},
	{color=red, only marks, mark=*, mark size=0.1pt, line width=0.25pt},
},
ylabel={},
}, 
ci graph style/.style={
xlabel style={yshift=1mm},
ylabel style={yshift=-1.5mm},
yticklabel style={xshift=1mm},
mark repeat = 1,
mark phase = 0,
ymin=1e-3,
ymax=100,
ytick = {100, 50, 10, 1, 0.1, 0.01, 1e-3, 1e-4},
yticklabels = {$0$, $50$, $90$, $99$, $99.9$, $99.99$, $99.999$, $99.9999$},
y dir=reverse,
},     
bp coeff style/.style={
scale only axis=true,
width=0.225*.9\linewidth,
height=0.225*.9\linewidth,
scale only axis,
xmin=-4.000,
xmax=4.000,
xlabel={$\ell${\color{white}$\aod$}},
ticklabel style={font=\footnotesize},
ymin=0.000, ymax=0.9,
ylabel={$c_\ell$},
xlabel style={font=\footnotesize},
ylabel style={font=\footnotesize},
major tick length=2pt
},
bp graph style/.style={        
scale only axis=true,
width=0.35*1.1\linewidth,
height=0.225*.9\linewidth,
scale only axis,
xmin=-3.14, xmax=3.14,
xlabel={$\aod${\color{white}$\ell$}},
ticklabel style={font=\footnotesize},
xtick={-3.14,-1.57,0.0,1.57,3.14},
xticklabels={$-\pi$,$-\tfrac{\pi}{2}$,$0$,$\tfrac{\pi}{2}$,$\pi$},
ymin=0.000, ymax=3,
ylabel={Beampattern},
xlabel style={font=\footnotesize}, ylabel style={font=\footnotesize},
major tick length=2pt
},
peb graph style/.style={        
width=0.66\linewidth,
scale only axis,
point meta min=-2.583,
point meta max=-0.300,
axis on top,
xmin=0.000,
xmax=12.000,
xlabel={x in meter},
y dir=reverse,
ymin=0.000,
ymax=8.000,
ylabel={y in meter},
ytick={7.0,6.0,...,0.0},
xtick={0.0,1.0,...,12.0},
yticklabels={$1$,$2$,$3$,$4$,$5$,$6$,$7$,$8$},
xlabel style={font=\scriptsize,yshift=0.125cm},
ylabel style={font=\scriptsize,yshift=-0.125cm},
ticklabel style={font=\scriptsize},
unit vector ratio*=1 1 1,
yticklabel pos=left,
major tick length=2pt,
colormap={mymap}{[1pt] rgb(0pt)=(1,1,1); rgb(1pt)=(0.858903,0.984776,0.839302); rgb(2pt)=(0.777958,0.94143,0.649487); rgb(3pt)=(0.755504,0.864264,0.463393); rgb(4pt)=(0.777509,0.754439,0.310168); rgb(5pt)=(0.820314,0.619497,0.21003); rgb(6pt)=(0.854796,0.471879,0.170327); rgb(7pt)=(0.851327,0.326629,0.183322); rgb(8pt)=(0.784671,0.198575,0.225774); rgb(9pt)=(0.637629,0.0993149,0.259577); rgb(10pt)=(0.400067,0.0343393,0.229819); rgb(11pt)=(0,0,0)},
colorbar style={ylabel={Position Error Bound in centimeter (logscale)}, ytick={-0.4,-0.82,...,-2.92}, yticklabels={$39.8$, $15.1$, $5.8$, $2.2$, $0.8$, $0.3$},ylabel style={yshift=0.5mm,font=\scriptsize,scale=0.8},width=2.0mm,xshift=-4.25mm,ticklabel style={font=\scriptsize},major tick length=0pt}, 
colormap access=piecewise constant
},
peb ellipses/.style={color=white, line width=0.4pt, forget plot}
}

%% file: base/tikzsymbols.tex
\tikzset{naming/.style={align=center,font=\small}}
\tikzset{antenna/.style={insert path={-- coordinate (ant#1) ++(0,0.25) -- +(135:0.25) + (0,0) -- +(45:0.25)}}}
\tikzset{station/.style={naming,draw,shape=dart,shape border rotate=90, minimum width=10mm, minimum height=10mm,outer sep=0pt,inner sep=3pt}}
\tikzset{mobile/.style={naming,draw,shape=rectangle,minimum width=12mm,minimum height=6mm, outer sep=0pt,inner sep=3pt}}
\tikzset{radiation/.style={{decorate,decoration={expanding waves,angle=90,segment length=4pt}}}}

\tikzset{
  pobl/.style={
    inner sep=0pt, outer sep=0pt, fill=#1,
  },
  pobl gron/.style n args={2}{
    pobl=#1, rounded corners=#2,
  },
  pics/person/.style n args={3}{
    code={
      \node (-corff) [pobl=#1, minimum width=.25*#2, minimum height=.375*#2, rotate=#3, pic actions] {};
      \node (-pen) [minimum width=.3*#2, circle, pobl=#1, outer sep=.01*#2, anchor=south, rotate=#3, pic actions] at (-corff.north) {};
      \node (-coes dde) [pobl gron={#1}{1pt}, anchor=north west, minimum width=.12125*#2, minimum height=.25*#2, rotate=#3, pic actions] at (-corff.south west) {};
      \node [pobl=#1, anchor=north, minimum width=.12125*#2, minimum height=.15*#2, rotate=#3, pic actions] at (-coes dde.north) {};
      \node (-coes chwith) [pobl gron={#1}{1pt}, anchor=north east, minimum width=.12125*#2, minimum height=.25*#2, rotate=#3, pic actions] at (-corff.south east) {};
      \node [pobl=#1, anchor=north, minimum width=.12125*#2, minimum height=.15*#2, rotate=#3, pic actions] at (-coes chwith.north) {};
      \node (-braich dde) [pobl gron={#1}{.75pt}, minimum width=.075*#2, minimum height=.325*#2, outer sep=.0064*#2, anchor=north west, rotate=#3, pic actions] at (-corff.north east)  {};
      \node [pobl=#1, minimum width=.05*#2, minimum height=.2*#2, outer sep=.0064*#2, anchor=north west, rotate=#3, pic actions] at (-corff.north east) {};
      \node (-braich chwith) [pobl gron={#1}{.75pt}, minimum width=.075*#2, minimum height=.325*#2, outer sep=.0064*#2, anchor=north east, rotate=#3, pic actions] at (-corff.north west) {};
      \node [pobl=#1, minimum width=.0375*#2, minimum height=.2*#2, outer sep=.0064*#2, anchor=north east, rotate=#3, pic actions] at (-corff.north west) {};
      \node (-fit person) [fit={(-pen.north) (-braich dde.east) (-coes chwith.south) (-braich chwith.west)}] {};
    },
  },
  pics/SBS/.style={code={
      \begin{scope}[local bounding box=#1]
      \fill [pic actions/.try] (-1,0) -- (-1/2,3) -- (1/2, 3) -- (1,0) -- cycle;
      \fill [pic actions/.try] (-1/16,2) rectangle (1/16,4);
      \fill [pic actions/.try] (0,4) circle [radius=1/4];
      \foreach \i in {-1,1}
        \fill [shift=(90:4), xscale=\i]
          \foreach \r in {1,3/2,2}{
            (-45:\r) arc (-45:45:\r) -- (45:\r-1/10)
            arc(45:-45:\r-1/10) -- cycle
          };
       \end{scope}
  }},
}

%% file: sections/abstract.tex
This paper addresses the challenge of achieving robust and reliable positioning of a radio device carried by an agent, in scenarios where direct \ac{los} radio links are obstructed by the agent. 
We propose a Bayesian estimation algorithm that integrates active measurements between the radio device and anchors with passive measurements in-between anchors reflecting off the agent. 
A geometry-based scattering measurement model is introduced for multi-sensor structures, and multiple object-related measurements are incorporated to formulate an extended object \ac{pda} algorithm, where the agent that blocks, scatters and attenuates radio signals is modeled as an \ac{eo}.
The proposed approach significantly improves the accuracy during and after obstructed LOS conditions, outperforming the conventional \ac{pda} (which is based on the point-target-assumption) and methods relying solely on active measurements. 

%% file: sections/introduction.tex
Localization and sensing have witnessed significant advancements in recent years.  
Integrating radar sensing with radio localization enables simultaneous positioning and tracking, crucial for applications like autonomous driving, keyless access system, and human activity recognition \cite{WitrisalSPM2016,LeitMeyHlaWitTufWin:TWC2019,Leitinger2023,Venus2024}. 
Additionally, multi-sensor frameworks enhance accuracy and reliability by leveraging spatial diversity and sensor fusion in dynamic environments\cite{VenusRadar2021,Venus2024,VenLeiTerMeyWit:TWC2024}.

\Ac{eot} addresses scenarios where objects, such as human bodies, generate multiple scattering paths due to their physical extent. Unlike traditional point-source models, \ac{eot} accounts for an object's spatial dimensions, offering a more accurate representation. Previous work includes modeling with ellipses, rectangles, star-convex shapes, and random matrices, effectively capturing the spatial distribution of scattering points, and utilizing probabilistic frameworks for state estimation \cite{Koch2008, Granstroem2014,Baum2014}. \Ac{pda} is a Bayesian approach used in target tracking to address measurement origin uncertainty \cite{BarShalomTCS2009}.
Conventional \ac{pda} \cite{BarShalomTCS2009} uses the ``point-target-assumption'' that disregards the extended nature of objects, leading to a model mismatch in scenarios where multiple measurements arise from spatially distributed scattering points\cite{MeyerTSP2021,Wielandner2023,WieVenWilWitLei:Fusion2024}. This limitation necessitates incorporating multiple object-related measurements into the data association process to enhance localization performance. 
Furthermore, when a radio device is carried by an agent (e.g. a person or a robot)\footnote{In this paper, the agent that can block, scatter and attenuate radio signals is referred to as \ac{eo}.}, \ac{los} radio links can be obstructed by the agent during certain time, which significantly deteriorates the radio localization performance. 
\begin{figure}[t]
	\centering
	\includegraphics[width=9cm]{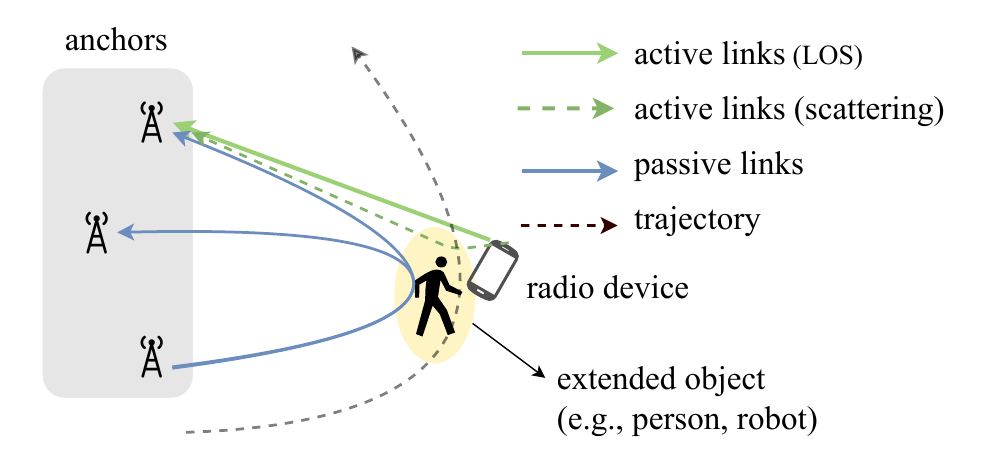}
	\vspace{-5mm}
	\caption{A radio device carried by a person, modeled as an extended object, moves along a trajectory.}
	\label{fig:concept}
	\vspace{-6mm}
\end{figure}

This paper presents a radio localization approach for extended object tracking in \ac{olos} scenarios. We propose a Bayesian estimation algorithm which fuses active measurements between a radio device and multiple anchor nodes (fixed radio transceivers with known position) as well as passive measurements in-between anchors reflected off the EO as illustrated in Fig.~\ref{fig:concept}. The algorithm estimates both the radio device's position and the object's extent using position-related information from LOS and scattering components. 
An efficient geometry-based scattering model is proposed to overcome the computational complexity of the ideal scattering model, while still allowing to fuse the scattering information from multiple anchors to jointly estimate the object's extent. Additionally, an extended object probabilistic data association (EOPDA) algorithm addresses the limitation of the point assumption PDA, improving positioning accuracy.

%% file: sections/signal_model.tex
\begin{figure}[t]
	\centering
	\captionsetup[subfloat]{captionskip=-2mm}
	\subfloat[\label{fig:Modelideal}]{\includegraphics{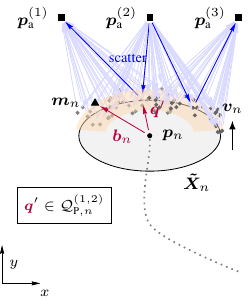}}\vspace{0mm}
	\subfloat[\label{fig:ModelEO}]{\includegraphics{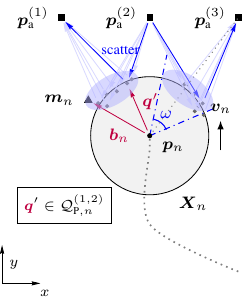}}\vspace{0mm}
	\caption{Scattering models of the extended object for passive measurements. (a) an ideal scattering model where scatterers are distributed within a sector on the  EO's surface, (b) a simplified geometry-based scattering model exploiting the geometric relation of scatterers and the EO.}
	\label{fig:Modelcompare}
	\vspace{-6mm}
\end{figure}

At each time step $n$, a radio device at position $\bm{m}_n$ transmits a signal, and each anchor $j \rmv\rmv \in \rmv\rmv \{1,\s... \s,J\}$ at position $\bm{p}_{\text{a}}^{(j)} = [p_{\text{ax}}^{(j)} \; p_{\text{ay}}^{(j)}]^\text{T}$ acts as a receiver, capturing active measurements. Synchronously, pairs of anchors $(j,j')$ exchange signals, capturing passive measurements from the \ac{eo}.
The \ac{eo}, centered at position
$\bm{p}_n$, is rigidly coupled to the radio device. The gap between the device and the EO's center is described by the bias $\bm{b}_n$. 
An example is shown in Fig.~\ref{fig:Modelideal}.
We assume scatterers are primarily distributed on the EO's surface, with the corresponding sector treated as a scattering volume\cite{Hoher2022}.
We denote the scattering volume for active and passive measurements as $\vm{Q}_{\text{A},n}^{(j)}$ and $\vm{Q}_{\text{P},n}^{(j,j')}$ for received anchor $j$ at time $n$.
Each point-source scatterer is denoted by its position $\vm{q} \in \vm{Q}_{\text{A},n}^{(j)}$ and $\vm{q'} \in \vm{Q}_{\text{P},n}^{(j,j')}$, respectively\cite{WildingEUCAP2022}.
\subsection{Active Radio Signal}
At each time $n$, a radio signal is transmitted from the radio device and received at anchor $j$. The complex baseband signal from anchor $j$ is modeled as
\vspace*{-1mm}
\begin{align}
	r_{\text{A}, n}^{(j)}(t) &= \alpha_{n}^{(j)}s(t-\tau_{n}^{(j)})\nn \\
	&\hspace*{7mm}+\sum_{\vm{q} \in \vm{Q}_{\text{A},n}^{(j)}} \alpha_{\vm{q},n}^{(j)}s(t-\tau_{\vm{q},n}^{(j)})+ w_n^{(j)}(t)
	\label{equ:rx_active}\\[-8mm]\nn
\end{align}
where $\alpha_{n}^{(j)}$ and $\tau_{n}^{(j)}$ are the complex amplitude and delay of the LOS component from active measurements. 
The complex amplitude and delay of the scatter component are denoted as $\alpha_{\vm{q},n}^{(j)}$ and $\tau_{\vm{q},n}^{(j)}$
The second term $w_n^{(j)}(t)$ accounts for measurement noise modeled as \ac{awgn} with double-sided power spectral density $N_0/2$. 
\subsection{Passive Radio Signal}
At each time step $n$, a radio signal is transmitted from anchor $j'$ and received at anchor $j$. The complex baseband signal from anchor $j$ is modeled as
\vspace*{-1mm}
\begin{align}
	r_{\text{P}, n}^{(j,j')}(t) = \sum_{\vm{q'} \in \vm{Q}_{\text{P},n}^{(j,j')}}\alpha_{\vm{q'},{n}}^{(j,j')}s(t-\tau_{\vm{q'},{n}}^{(j,j')}) + w_{n}^{(j,j')}(t)
	\label{equ:rx_passive}\\[-7mm]\nn
\end{align}
where $ \alpha_{\vm{q'},{n}}^{(j,j')}$ and $\tau_{\vm{q'},{n}}^{(j,j')}$ are the complex amplitude and delay of the scatter component from $\vm{Q}_{\text{P},n}^{(j,j')}$.
\subsection{Signal Parameter Estimation}\label{eq:CEDA}
Measurements which are extracted using a channel estimation algorithm \cite{Venus2024} from active radio signals are called active measurements $\V{z}_{\text{A},n}$, while measurements which are extracted from passive radio signals are called passive measurements $\V{z}_{\text{P},n}$.
We define the vectors $\bm{z}_{\text{A},n} = [{\bm{z}_{\text{A},n}^{(1)}}^\text{T} \cdots {\bm{z}_{\text{A},n}^{(J)}}^{\text{T}}]^\text{T}$ and $\bm{z}_{\text{P},n} = [{\bm{z}_{\text{P},n}^{(1,1)}}^{\text{T}} \cdots {\bm{z}_{\text{P},n}^{(J,J)}}^{\text{T}}]^\text{T}$ for the measurement vectors per time $n$. 
Take the passive case for example, we define $\V{z}^{(j,j')}_{\text{P},n}= [\V{z}^{(j,j')}_{\text{P},n,1},\dots\,, \V{z}^{(j,j')}_{\text{P},n, M_{\text{P}, n}^{(j,j')}}]$ with $M_{\text{P},n}^{(j,j')}$ being the number of passive measurements. Each passive measurement $\V{z}^{(j,j')}_{\text{P},n,l}= [z_{\text{P},\text{d},n,l}^{(j,j')} ~ z_{\text{P},\text{u},n,l}^{(j,j')}]^\text{T}$, $l \in  \{1,\,\dots\,M_{\text{P},n}^{(j,j')}\}$ contains a distance measurement $z_{\text{P},\text{d},n,l}^{(j,j')} \in [0, d_\text{max}]$ and a normalized amplitude measurement $z_{\text{P}, \text{u},n,l}^{(j,j')} \in [\gamma, \infty)$.

%% file: sections/system_model.tex
In this section, we formulate a Bayesian method which fuses multiple object-related measurements from both active and passive measurement data. We jointly estimate the kinematic states as well as the extent states of the \ac{eo} to address the extended object tracking problem.

At time $n$, the radio device and the extended object are characterized by the kinematic state, bias state and extent state. 
The kinematic state  $\bm{x}_n = [\bm{p}_n ^\text{T}\; \bm{v}_n^\text{T}]^\text{T}$ consists of the position of the EO's center $\bm{p}_n = [p_{\text{x}\s n}\; p_{\text{y}\s n}]^\text{T}$ and the velocity $\bm{v}_n = [v_{\text{x}\s n}\; v_{\text{y}\s n}]^\text{T}$. 
The bias describes the offset between the EO center and the radio device, defined as $\bm{b}_n = [b_{\rho\s n}\; b_{\phi\s n}]^\text{T}$, where $b_{\rho}$ is the distance between the EO center and the radio device, and $b_{\phi}$ is the orientation relative to the x-axis of the EO coordinate system. 
A geometry-based scattering model is proposed to approximate the extent of the EO, as illustrated in Fig.~\ref{fig:ModelEO}. 
The EO is approximated as a circle, while the scattering volume is modeled as an ellipse, referred to as scattering ellipse.
The extent state ${\bm{X}_n} =[r_n\; w_n]^\text{T}$, where $r_n$ denotes the circle's radius, and $w_n$ represents the  semi-minor axis of the scattering ellipses for all anchors. 
For simplicity, we define the augmented extended object state as $\bm{y}_n = [\bm{x}_n ^\text{T}\; \bm{b}_n^\text{T}\;\bm{X}_n ^\text{T} ]^\text{T}$.
The state estimate of $\bm{y}_n$ is obtained by calculating the \ac{mmse} estimator $\hat{\bm{y}}^\text{MMSE}_{n} \,\triangleq \int \rmv \bm{y}_{n} \, f(\bm{y}_{n} | \V{z}_{\text{A},1:n},\V{z}_{\text{P},1:n} )\, \mathrm{d}\bm{y}_{n} \,$.
The estimation process involves marginalizing the joint posterior distribution, as detailed in Sec.\ref{sec:joint posterior}.
\subsection{LOS Measurement Model}
The position of the radio device is given as
\vspace*{-1mm}
\begin{align}
	\bm{m}_{n} = \bm{p}_n+
	\begin{bmatrix}
		b_{\rho \s n} \cos(b_{\phi \s n}) \\ b_{\rho \s n} \sin(b_{\phi \s n}) 
	\end{bmatrix} \, .
	\label{equ:pos_mobile_agent}\\[-7mm]\nn
\end{align}
The \ac{lhf} of an LOS path is given by
\begin{equation} 
	f_{\text{LOS}}(\V{z}^{(j)}_{\text{A},n,l}|\bm{x}_n,\bm{b}_n) = f_{\text{N}}(z_{\text{A},\text{d},n,l}^{(j)}; h_\text{LOS}(\bm{m}_n,\bm{p}_{\text{a}}^{(j)}), \sigma_{\text{d}}^{2} (z_{\text{A},\text{u},n,l}^{(j)})) 
	\label{equ: measLikelihood_active_LOS}
\end{equation}
where $f_{\text{N}}(x;\mu,\sigma^2)$ is the Gaussian PDF, with mean $h_\text{LOS}(\bm{m}_n,\bm{p}_{\text{a}}^{(j)}) = \|\vm{m}_n - \vm{p}_\text{a}^{(j)}\|$ being the LOS distance and variance $\sigma_{\text{d}}^{2} (z^{(j)}_{\text{A},\text{u},n,l})$. The variance is determined from the Fisher information given by
$ \sigma_{\text{d}}^{2} (z^{(j)}_{\text{A},\text{u},n,l}) =   c^2 / ( 8\,  \pi^2 \, \beta_\text{bw}^2 \, (z^{(j)}_{\text{A},\text{u},n,l})^2)$, where $\beta_\text{bw}$ is the root mean squared bandwidth \cite{WitrisalJWCOML2016,LeitingerJSAC2015}, and $(z^{(j)}_{\text{A},\text{u},n,l})^2$ corresponds to the SNR.
\subsection{Scattering Measurement Model}\label{sec:scatterModel}
The scattering \ac{lhf} conditioned on $\bm{x}_n$ and $\bm{X}_n$ is a convolution of the noise distribution and the scattering distribution \cite{MeyerTSP2021}. 
The \ac{lhf} of an individual scattering measurement is obtained by integrating out the scattering variables.
For the passive measurements $\V{z}^{(j,j')}_{\text{P}, n,l}$ it is given as 
\begin{align}
	\vspace{-3mm}
	f_{\text{P}}(\V{z}^{(j,j')}_{\text{P}, n,l}|\bm{x}_n, {\bm{X}}_n) 
	=\int f(\V{z}^{(j,j')}_{\text{P},n,l}|\bm{x}_{n}, \bm{q}')f(\bm{q}'|\bm{X}_{n}) d\bm{q}'
	\label{equ:scatterMeasLikelihood}\\[-6mm]\nn
\end{align}

The proposed geometry-based scattering model approximates the scattering distribution in (\ref{equ:scatterMeasLikelihood}) as follows.
For each received anchor $j$, the center $\bm{\chi}_n^{(j)}$ of the scattering ellipse is represented as 
\vspace{-2mm}
\begin{align}
	\bm{\chi}_{n}^{(j)} = \bm{p}_n+
	\begin{bmatrix}
		r_n \cos(\phi_n^{(j)}) \\ r_n \sin(\phi_n^{(j)})
	\end{bmatrix}
	\label{equ:intersection}\\[-6mm]\nn
\end{align}
where $\phi_n^{(j)}$ is the angle between the x-axis of the \ac{eo} coordinate system and the line from $\bm{p}_n$ to $\bm{p}_{\text{a}}^{(j)}$ given as $\phi_n^{(j)}=\text{atan2}(\frac{p_{\text{y}\s n} -{p}_{\text{a}\s \text{y}}^{(j)}}{p_{\text{x}\s n} -{p}_{\text{a}\s \text{x}}^{(j)}})$.
While the unified semi-minor axis $w_{n}$ of all scattering ellipses (contained in $\bm{X}_n$) is jointly estimated, the semi-major axis  $l_{n}^{(j)}$ is determined by the
opening angle $\omega$, symmetric to the line connecting $\bm{p}_n$ to $\bm{p}_{\text{a}}^{(j)}$ (see Fig.~\ref{fig:ModelEO}), capturing scatterers from each anchor $j$. The orientation $\theta_{n}^{(j)}$ of the scattering ellipse follows the circle's tangent direction.
The measurement covariance can be represented as $\bm{R}_n^{(j)} = \bm{A}_n^{(j)}\bm{E}_n^{(j)}{\bm{A}_n^{(j)}}^\text{T}$, while $\bm{E}_n^{(j)} \in \mathbb{R}^{2 \times 2}$ is a symmetric, positive semidefinite matrix that describes the 2-D scattering ellipse, and $\bm{A}_n^{(j)}$ is the rotation matrix related to $\theta_{n}^{(j)}$.
We denote the larger eigenvalue of $\bm{E}_n^{(j)}$ as $e_1$ and the smaller ones as $e_2$.
It is assumed that the square root of these eigenvalues is proportional to the volume's semi-axis \cite{Hoher2022}. This leads to
\vspace{-1mm}
\begin{align}
	l_n^{(j)} = 2\sqrt{e_{1}} ~~~\text{and}~~~ 
	w_n = 2\sqrt{e_{2}}\\[-6mm]\nn
\end{align}
Take the passive case for example, the measurement covariance $\bm{R}_n^{(j)}$ is used to decide the covariance matrix of a Gaussian PDF \footnote{It is shown in \cite{Feldmann2011} that for an elliptically shaped object the uniform distribution can be approximated by a Gaussian distribution.} that models the scattering distribution due to the geometric shape as
$f(\bm{\zeta}|\bm{R}_n^{(j)})\,\triangleq f_{\text{N}}(\bm{\zeta}; \bm{0},\bm{R}_n^{(j)})$.
The \ac{lhf} of an individual scattering measurement is derived as
\vspace{-2mm}
\begin{align}
	&f_{\text{P(geo)}}(\V{z}^{(j,j')}_{\text{P},n,l}|\bm{x}_n, \bm{X}_n) \nn\\
	&=\int f(\V{z}^{(j,j')}_{\text{P},n,l}|\bm{\chi}_{n}^{(j)}, \bm{\zeta})f(\bm{\zeta}|\bm{R}_n^{(j)}) d\bm{\zeta} \nn \\
	& = f_{\text{N}}(z_{\text{P}, \text{d},n,l}^{(j,j')}; {h}_\text{P}(\bm{\chi}_{n}^{(j)},\bm{p}_{\text{a}}^{(j')},\bm{p}_{\text{a}}^{(j)}), \sigma_{\text{d}}^{2} (z_{\text{P},\text{u},n,l}^{(j,j')})+\sigma_{\lambda}^{2})
	\label{equ: measLikelihood_passive}\\[-6mm]\nn
\end{align}
where ${h}_\text{P}(\bm{\chi}_{n}^{(j)}, \bm{p}_{\text{a}}^{(j')}, \bm{p}_{\text{a}}^{(j)}) = \|\bm{\chi}_{n}^{(j)}- \bm{p}_\text{a}^{(j')}\|+ \|\bm{\chi}_{n}^{(j)}- \bm{p}_\text{a}^{(j)}\|$. 
The unscented transformation (UT)\cite{Wan2000} is applied to transform the propagation uncertainty of the scattering distribution $f(\bm{\zeta}|\bm{R}_n^{(j)})$ from position domain to delay domain.
The variance $\sigma_{\lambda}^{2}$ represents the dispersion of the sigma points transformed through the nonlinear function $h_\text{P}(\cdot)$.
\subsection{Data Association Uncertainty}
For each anchor $j$, the measurements $\V{z}_{\text{A},n}^{(j)}$ and $\V{z}_{\text{P},n}^{(j,j')}$ are subject to data association uncertainty. Specifically, it is unknown whether a given measurement corresponds to the \ac{los} path, the extended object, or is a result of clutter.
To address this uncertainty, the association variables $a_{\text{A},n,l}^{(j)} \in \{0,1\}$ and $a_{\text{P},n,l}^{(j,j')} \in \{0,1\}$   are introduced, where a value of $1$ indicates that a measurement is LOS or object-related, while a value of $0$ indicates otherwise.
Object-related and clutter measurements follow Poisson distribution with means $\mu_{\text{m}}$ and $\mu_{\text{c}}$. Clutter measurements are independent and uniformly distributed according to $f_c(\V{z}_{\text{A},n,l}^{(j)})$ and $f_c(\V{z}_{\text{P},n,l}^{(j,j')})$. 
\subsection{Joint Posterior PDF}\label{sec:joint posterior}
It is assumed that the state $\bm{y}_n$ evolves over time $n$ as an independent first-order Markov process. Therefore, the joint state transition \ac{pdf} can be represented as 
\vspace{-1mm}
\begin{align}
	f(\bm{y}_n | \bm{y}_{n-1}) = f(\bm{x}_{n}|\bm{x}_{n-1})
	f(\bm{b}_{n}|\bm{b}_{n-1}) f(\bm{X}_{n}|\bm{X}_{n-1})
	\label{equ: state_transition_model}\\[-6mm]\nn
\end{align}
where $f({\bm{x}}_n|{\bm{x}}_{n-1})$, $f(\bm{b}_n|\bm{b}_{n-1})$ and $f(\bm{X}_n|\bm{X}_{n-1})$ are the state transition \acp{pdf} of the agent motion, bias and the extent parameters.

\begin{figure}[t]
	\centering
	\includegraphics{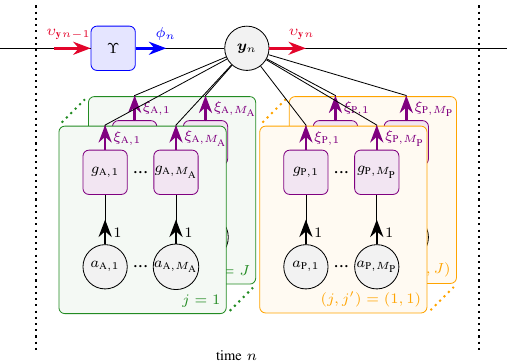}
	\caption{Factor graph representing the factorization of the joint posterior \ac{pdf} in \eqref{equ: joint posterior} and the messages according to the SPA.
		The following short notations are used: $M_{\text{A}} \triangleq M_{\text{A},n}^{(j)}$, $M_{\text{P}} \triangleq M_{\text{P},n}^{(j)}$, $a_{\text{A},l} \triangleq a_{\text{A},n,l}^{(j)}$, $a_{\text{P},l} \triangleq a_{\text{P},n,l}^{(j,j')}$, ${g}_{\text{A},l} \triangleq {g}_{\text{A},n,l}^{(j)}$, ${g}_{\text{P},l} \triangleq {g}_{\text{P},n,l}^{(j,j')}$, $\xi_{\text{A},l}\triangleq \xi_{\text{A},n,l}^{(j)}$, $\xi_{\text{P},l}\triangleq \xi_{\text{P},n,l}^{(j,j')}$.}
	\label{fig:factorGraph}
	\vspace*{-2mm}
\end{figure}
We assume that the measurements $\V{z}_{\text{A},n}^{(j)}$ and $\V{z}_{\text{P},n}^{(j,j')}$ are observed and thus fixed. 
According to the Bayes's rule and the related  independence assumptions, the joint posterior PDF of all estimated states for time $n$ and all $J$ anchors can be derived as
\vspace{-2mm}
\begin{align}
	&f(\bm{y}_{0:n},\bm{a}_{\text{A},{1:n}}, \bm{a}_{\text{P},{1:n}} | \V{z}_{\text{A},{1:n}}, \V{z}_{\text{P},{1:n}})\nn\\[-1mm] 
	&\propto f(\bm{y}_0) \prod_{n'=1}^{n} \Upsilon(\bm{y}_{n'}|\bm{y}_{n'-1}) \times \prod_{j=1}^{J} \prod_{l=1}^{M_{\text{A},n'}^{(j)}} {g}_{\text{A}}(\bm{z}_{\text{A},n',l}^{(j)}|\bm{y}_{n'}, a_{\text{A},n',l}^{(j)})\nn\\[-1mm] 
	&\hspace{6mm} \times \prod_{j'=1}^{J} \prod_{l=1}^{M_{\text{P},n'}^{(j,j')}} {g}_{\text{P}}(\bm{z}_{\text{P},n',l}^{(j,j')}|\bm{y}_{n'}, a_{\text{P},n',l}^{(j,j')})
	\label{equ: joint posterior}\\[-7mm]\nn
\end{align}
where $\Upsilon(\bm{y}_n|\bm{y}_{n-1}) \,\triangleq f(\bm{y}_n|\bm{y}_{n-1})$.
Fig.~\ref{fig:factorGraph} is the factor graph that represents the factorization of (\ref{equ: joint posterior}).
The pseudo-likelihood function for the passive case is represented as
\vspace{-2mm}
\begin{align}
	{g}_{\text{P}}(\V{z}_{\text{P},n,l}^{(j,j')}|\bm{y}_n, a_{\text{P},n,l}^{(j,j')}) 
	= \begin{cases}
		\frac{\mu_m f_\text{P}(\V{z}_{\text{P},n,l}^{(j,j')}|\bm{x}_n,  \bm{X}_n)}{\mu_c f_c(\V{z}_{\text{P},n,l}^{(j,j')})}, &\text{$a_{\text{P},n,l}^{(j,j')} = 1$}\\ 
		1, & \text{$a_{\text{P},n,l}^{(j,j')} = 0$}
	\end{cases}
	\label{equ: passive pseudo-measurement likelihood}\\[-7mm]\nn
\end{align}
To estimate the states, marginalization of the joint posterior is performed by message passing on the factor graph in Fig.~\ref{fig:factorGraph} using the \ac{spa}\cite{KschischangTIT2001} and a particle-based implementation similar to \cite{Venus2024}.

%% file: sections/ideal.tex
For comparison, we also approximate the scattering distribution (\ref{equ:scatterMeasLikelihood}) based on the ideal scattering model in  Fig.~\ref{fig:Modelideal}, where the EO is represented as an ellipse with scatterers distributed within a sector on its surface.
The extent state $\bm{\tilde{X}}_n =[a_n\; b_n\; w_n]^\text{T}$, where $a_n$, $b_n$ denote the semi-major axis and the semi-minor axis of the EO, respectively, and $w_n$ represents the width of the sector. 
The scattering \ac{lhf} is approximated by a Monte Carlo technique sampling within the sector to evaluate the integral\vspace*{-2mm}
\begin{align}
	&f_{\text{P(idl)}}(\V{z}^{(j,j')}_{\text{P}, n,l}|\bm{x}_n,{\bm{\tilde{X}}}_n)  \\
	& \approx \frac{1}{I'} \sum_{i=1 }^{I'} f_{\text{N}}(z_{\text{P},\text{d},n,l}^{(j,j')}; \bar{h}_\text{P}(\bm{x}_n, \bm{q}'_i,\bm{p}_{\text{a}}^{(j')}, \bm{p}_{\text{a}}^{(j)}), \sigma_{\text{d}}^{2} (z_{\text{P},\text{u},n,l}^{(j,j')})) \nn
	\label{equ: measLikelihood_ideal}\\[-7mm]\nn
\end{align}
where $\bar{h}_\text{P}(\bm{x}_{n}, \bm{q}'_i, \bm{p}_{\text{a}}^{(j')}, \bm{p}_{\text{a}}^{(j)}) = \|(\bm{p}_{n}+\bm{q}'_i)- \bm{p}_\text{a}^{(j')}\|+ \|(\bm{p}_{n}+\bm{q}'_i)- \bm{p}_\text{a}^{(j)}\|$, 
$\bm{q}'_i$ is the random sample generated in the sector,
and $I'$ is the number of the samples used per received anchor.

%% file: sections/results.tex

\subsection{Simulation Setup} \label{sec:simulationSetting}
\begin{figure}[t]
	\vspace{-6mm}
	\centering
	\setlength{\abovecaptionskip}{0pt}
	\setlength{\belowcaptionskip}{0pt}
	
	\setlength{\figurewidth}{0.28\textwidth}
	\setlength{\figureheight}{0.28\textwidth}
	
	\includegraphics{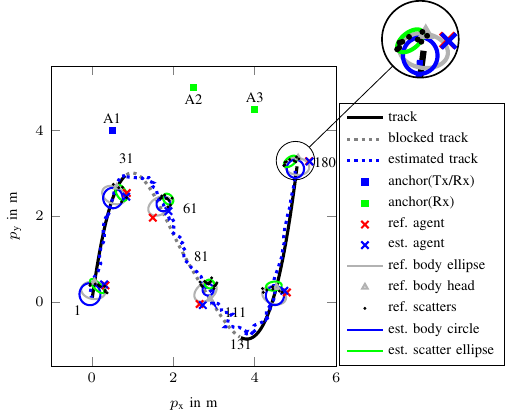}

	\caption{Graphical representation of the synthetic trajectory and one realization of AP-EOPDA(geo) method. The scatters are generated with respect to one received anchor at position $(4,4.5)$ for each time.
	}\label{fig:simulationScenario}
	\vspace{-6mm}
\end{figure}

The proposed algorithm is evaluated using synthetic measurements generated according to the scenario presented in Fig.~\ref{fig:simulationScenario} and the radio signal models introduced in Section \ref{sec:signal_model}. 
The EO moves along a smooth trajectory featuring two direction changes.
A radio device is rigidly coupled to the EO with $b_{\rho} = 0.32\,\mathrm{m}$ and $b_{\phi} = -\pi/3$. 
The generative model follows the ideal scattering model with $\bm{\tilde{X}}$, where $a = 0.3\,\mathrm{m}$, $b = 0.2\,\mathrm{m}$, and $w = 0.1\,\mathrm{m}$.
The opening angle is set to  $\omega = 2\pi/3$.
The mean number of scattering measurements and clutter are  $\mu_m=5$ and $\mu_c=5$, respectively.
The normalized amplitudes are set to $30$ dB at a $1\,\mathrm{m}$ \ac{los} distance and follow free-space pathloss.
The object is observed at $180$ discrete time steps at a constant observation rate of $\Delta T = 100\,\mathrm{ms}$.
Active measurements are entirely missed for anchors [A1, A2, A3] during time steps $[31, 60]$ and $[111, 130]$, for [A1, A2] during $[61, 80]$, and for [A2] during $[81, 110]$, while passive paths from all anchors remain available throughout the trajectory.

The state transition \ac{pdf} of the kinematic state  $f({\bm{x}}_n|{\bm{x}}_{n-1})$ is described by a linear, constant velocity and stochastic acceleration model\cite[p.~273]{BarShalom2002EstimationTracking}, given as ${\RV{x}}_n = \bm{A}\, {\RV{x}}_{n\minus 1} + \bm{B}\, \RV{w}_{n}$. 
The acceleration process $\RV{w}_n$ is i.i.d. across $n$, zero mean, and Gaussian with covariance matrix ${\sigma_{\text{a}}^2}\, \bm{I}_2$, with ${\sigma_{\text{a}}}$
being the acceleration standard deviation, and $\bm{A} \in \mathbb{R}^{\text{4x4}}$ and $\bm{B} \in \mathbb{R}^{\text{4x2}}$ are defined according to \cite[p.~273]{BarShalom2002EstimationTracking}.
Furthermore, the state transition PDF of the bias state is factorized as $f(\bm{b}_n|\bm{b}_{n-1})= f({b}_{\rho \s n}|{{b}}_{\rho \s {n-1}})f({b}_{\phi \s n}|{b}_{\phi \s {n-1}})$.
The PDFs of ${b}_{\rho \s n}$ and ${b}_{\phi \s n}$ are ${b}_{\rho \s n}={b}_{\rho \s n-1} + {\varepsilon}_{\rho \s n}$ and ${b}_{\phi \s n}={b}_{\phi \s n-1} + {\varepsilon}_{\phi \s n}$, respectively. 
For the geometry-based scattering model, the state transition PDF of the extent state is factorized as $f(\bm{X}_n|\bm{X}_{n-1})= f({r}_{n}|{{r}}_{{n-1}})f({w}_{n}|{w}_{{n-1}})$.
The PDFs of ${r}_{n}$ and ${w}_{n}$ are ${r}_{n}={r}_{n-1} + {\varepsilon}_{r \s n}$ and ${w}_{n}={w}_{n-1} + {\varepsilon}_{w \s n}$, respectively.
While the noise ${\varepsilon}_{\rho \s n}$, ${\varepsilon}_{\phi \s n}$, ${\varepsilon}_{r\s n}$ and  ${\varepsilon}_{w\s n}$ are i.i.d. across $n$, zeros mean, Gaussian, with variances ${\sigma_{{\varepsilon}_{\rho}}^2}$, ${\sigma_{{\varepsilon}_{\phi}}^2}$,
${\sigma_{{\varepsilon}_{r}}^2}$ and ${\sigma_{{\varepsilon}_{w}}^2}$, respectively.
The number of particles is set to $I=5000$ for inference during the track, and the particles consist of all considered random variables.
The state-transition variances are set as ${\sigma_{\text{a}}} = 2\,\mathrm{m/s^2}$, ${\sigma_{\bm{\varepsilon}_{\rho}}}=0.1\,\mathrm{m}$, ${\sigma_{\bm{\varepsilon}_{\phi}}}=0.5\,\mathrm{rad}$, ${\sigma_{\bm{\varepsilon}_{r}}}=0.05\,\mathrm{m}$, and ${\sigma_{\bm{\varepsilon}_{w}}}=0.05\,\mathrm{m}$.

\subsection{Performance Evaluation}
To evaluate the proposed algorithm, we compare the PDA method under the point assumption, denoted as AP-PDA, with the PDA designed for the extended object, denoted as AP-EOPDA, and a method excluding passive measurements, denoted as A-EOPDA.
Additionally, we contrast the proposed geometry-based scattering model with the ideal scattering model, as mentioned in Sec.~\ref{sec:ideal}, under the AP-EOPDA method, denoted as AP-EOPDA(geo) and AP-EOPDA(idl) with $I'=50$.
The \ac{pcrlb} is provided as a performance baseline considering the dynamic model of the EO state \cite{Tichavsky1998}. The P-CRLB* assumes continuous LOS availability to all anchors throughout the trajectory, while the P-CRLB varies based on LOS blockages along the trajectory.

\begin{figure}[t]
	\captionsetup[subfloat]{captionskip=-2mm}
	\centering
	\setlength{\abovecaptionskip}{0pt}
	\setlength{\belowcaptionskip}{0pt}
	\subfloat[\label{fig:rmse_noCE_500ideal}]{\includegraphics{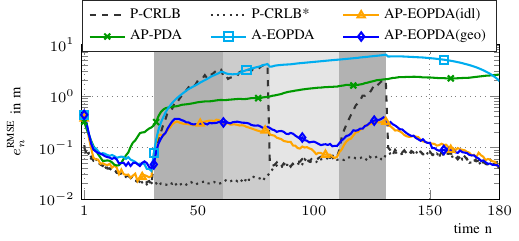}}\vspace{-0mm}
	\subfloat[\label{fig:cdf_noCE_500ideal}]{\includegraphics{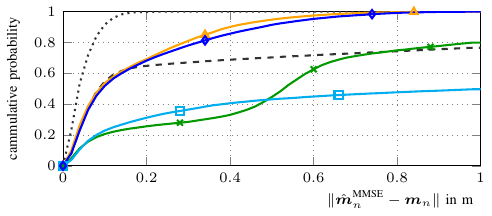}}
	\caption{Performance of different methods in synthetic measurements described in Fig.~\ref{fig:simulationScenario}, (a) is the RMSE of the estimated agent position, and (b) is the cumulative distribution of the RMSE based on numerical simulations. Different shades of grey represent different numbers of blocked anchors described in Sec.~\ref{sec:simulationSetting}.} 
	\label{fig:syn_noCE_500ideal}
	\vspace{-6mm}
\end{figure}

Fig.~\ref{fig:syn_noCE_500ideal} provides the results of a numerical simulation with $100$ runs. 
The \ac{rmse} of the estimated device's position is given by Fig.~\ref{fig:rmse_noCE_500ideal} and calculated by $e_{n}^{\text{RMSE}}~=~\sqrt{\E{\norm{\hat{\bm{m}}^{\text{MMSE}}_n -\bm{m}_n}{2}}}$.
Fig.~\ref{fig:cdf_noCE_500ideal} provides the cumulative probability of the position errors ${\|{\hat{\bm{m}}^{\text{MMSE}}_n -\bm{m}_n}\|}$ evaluated over the whole track.
Comparing A-EOPDA with AP-EOPDA, we find that the \ac{rmse} of the joint estimation (active \& passive) significantly outperforms that of the active-only estimation, particularly during and after the \ac{olos} time steps. 
AP-EOPDA(idl) precisely attains the P-CRLB before the OLOS situation and converges back to the P-CRLB afterward. In comparison, AP-EOPDA(geo) achieves a similar performance while requiring only half the execution time of the AP-EOPDA(idl) method, as shown in Table~\ref{tbl:execution_times}.
In contrast, AP-PDA diverges significantly to an incorrect position due to the inadequacy of the single-point assumption for extended object tracking.

\begin{table}[h]
	\renewcommand{\baselinestretch}{1}\small\normalsize
	\setlength{\tabcolsep}{3pt} 
	\renewcommand{\arraystretch}{1} 
	\centering
	\footnotesize
	\vspace{-3mm}
	\caption{Comparison of running time and averaged rmse values of different methods in investigated scenarios.}\label{tbl:execution_times}
	\begin{tabular}{ r c c c c} 
		\toprule
		\textbf{Models}  &  \textbf{avg. RMSE (m)} & \textbf{running time per step (s)} \\
		\midrule
		\multicolumn{1}{r}{AP-EOPDA(geo)} &0.18& 0.33 \\
		\multicolumn{1}{r}{AP-EOPDA(idl)} &0.16& 0.67 \\
		\bottomrule
	\end{tabular}
	\vspace{-2.5mm}
\end{table}

%% file: sections/conclusion.tex
This paper addresses the challenge of achieving robust  positioning of a radio device attached with an \ac{eo} when the \ac{los} between the device and anchors is obstructed by the \ac{eo}. 
We propose a joint estimation method that fuses both active and passive measurements from multiple anchors, introducing the probabilistic data association for extended object tracking. 
Results show that our proposed method significantly reduces the \ac{rmse} during and after the obstructed LOS, compared to methods using only active measurements or the point-assumption PDA. The passive measurements provide useful information for estimation, improving positioning accuracy during the full LOS blockage and minimizing outliers after the obstruction. 
Additionally, the geometry-based extended object model offers substantial computational efficiency, reducing the processing time by $50\,\%$ compared to the ideal scattering model, which is advantageous with an increasing anchor number. 
Future research will focus on validating the algorithm with real measurement data.

%% file: sections/acronyms.tex
 
 \acrodef{mimo}[MIMO]{multiple input multiple output}
 \acrodef{awgn}[AWGN]{additive white Gaussian noise}
 \acrodef{bw}[BW]{bandwidth}
 \acrodef{blt}[BLT]{bluetooth}
 \acrodef{cdf}[CDF]{cumulative distribution function}
 \acrodef{crlb}[CRLB]{Cram\'er-Rao lower bound}
 \acrodef{dmc}[DMC]{dense multipath component}
 \acrodef{dut}[DUT]{device under test}
 \acrodef{eo}[EO]{extended object}
 \acrodef{eirp}[EIRP]{equivalent isotropic radiated power}
 \acrodefplural{esl}[ESLs]{electronic shelf labels} 
 \acrodef{los}[LOS]{line-of-sight}
 \acrodef{mf}[MF]{matched filter}
 \acrodef{ml}[ML]{maximum likelihood}
 \acrodef{mpc}[MPC]{multipath component}
 \acrodef{nlos}[NLOS]{non-\ac{los}}
 \acrodef{eot}[EOT]{extended object tracking}
 \acrodef{pcb}[PCB]{printed circuit board}
 \acrodef{pdf}[PDF]{probability density function}
 \acrodef{reb}[REB]{ranging error bound}
 \acrodef{rss}[RSS]{received signal strength}
 \acrodef{smc}[SMC]{specular multipath component}
 \acrodef{snr}[SNR]{signal-to-noise-ratio}
 \acrodef{sinr}[SINR]{signal-to-interference-plus-noise-ratio}
 \acrodef{tdoa}[TDOA]{time difference of arrival}
 \acrodef{tka}[TKA]{trusted keyless access}
 \acrodef{toa}[TOA]{time-of-arrival}
 \acrodef{aoa}[AOA]{angle-of-arrival}
 \acrodef{uwb}[UWB]{ultra wide band}
 \acrodef{mie}[MIE]{method of interval estimation}
 \acrodef{mc}[MC]{Monte Carlo}
 \acrodef{mse}[MSE]{mean squared error}
 \acrodef{ci}[CI]{confidence interval}
 \acrodef{cl}[CL]{confidence level}
 \acrodef{pdp}[PDP]{power delay profile}
 \acrodef{dps}[DPS]{delay power spectrum}
 \acrodef{dm}[DM]{dense multipath}
 \acrodef{nlike}[NLIKE]{normalized likelihood}
 \acrodef{zzb}[ZZB]{Ziv-Zakai bound}
 \acrodef{ut}[UT]{unscented transform}
 \acrodef{glrt}[GLRT]{generalized likelihood ratio test}
 \acrodef{mse}[MSE]{mean squared error}
 \acrodef{rmse}[RMSE]{root mean squared error}
 \acrodef{nnlike}[NNLIKE]{normalized noise-free likelihood}
 \acrodef{stdv}[STDV]{standard deviation}
 \acrodef{rv}[RV]{random variable}
 \acrodef{bp}[BP]{belief propagation}
 \acrodef{pda}[PDA]{probabilistic data association}
 \acrodef{mp}[MP]{multipath}
 \acrodef{pmf}[PMF]{probability mass function}
 \acrodef{pdaf}[PDAF]{probabilistic data association filter}
 \acrodef{pdaai}[AIPDA]{amplitude-information \ac{pda}}
 \acrodef{olos}[OLOS]{obstructed line-of-sight}
 \acrodef{spa}[SPA]{sum-product algorithm}
 \acrodef{mmse}[MMSE]{minimum mean-square error}
 \acrodef{lhf}[LHF]{likelihood function}
 \acrodef{fa}[FA]{false alarm}
 \acrodef{ceda}[CEDA]{channel estimation and detection algorithm} 
 \acrodef{pcrlb}[P-CRLB]{posterior Cram\'er-Rao lower bound}
 \acrodef{slam}[SLAM]{simultaneous localization and mapping}
 \acrodef{mpslam}[MP-SLAM]{multipath-based SLAM}
 \acrodef{va}[VA]{virtual anchor}
 \acrodef{dnr}[DNR]{dense-to-noise ratio}
 \acrodef{aednn}[AE-DNN]{auto encoder deep neural network}   
 \acrodef{gpr}[GPR]{gaussian process regression}  
 \acrodef{ae}[AE]{auto encoder}

%% file: main.bbl
\begin{thebibliography}{10}
\providecommand{\url}[1]{#1}
\csname url@samestyle\endcsname
\providecommand{\newblock}{\relax}
\providecommand{\bibinfo}[2]{#2}
\providecommand{\BIBentrySTDinterwordspacing}{\spaceskip=0pt\relax}
\providecommand{\BIBentryALTinterwordstretchfactor}{4}
\providecommand{\BIBentryALTinterwordspacing}{\spaceskip=\fontdimen2\font plus
\BIBentryALTinterwordstretchfactor\fontdimen3\font minus
  \fontdimen4\font\relax}
\providecommand{\BIBforeignlanguage}[2]{{%
\expandafter\ifx\csname l@#1\endcsname\relax
\typeout{** WARNING: IEEEtran.bst: No hyphenation pattern has been}%
\typeout{** loaded for the language `#1'. Using the pattern for}%
\typeout{** the default language instead.}%
\else
\language=\csname l@#1\endcsname
\fi
#2}}
\providecommand{\BIBdecl}{\relax}
\BIBdecl

\bibitem{WitrisalSPM2016}
K.~Witrisal, P.~Meissner, E.~Leitinger, Y.~Shen, C.~Gustafson, F.~Tufvesson,
  K.~Haneda, D.~Dardari, A.~F. Molisch, A.~Conti, and M.~Z. Win,
  ``High-accuracy localization for assisted living: {5G} systems will turn
  multipath channels from foe to friend,'' \emph{{IEEE} Signal Process. Mag.},
  vol.~33, no.~2, pp. 59--70, Mar. 2016.

\bibitem{LeitMeyHlaWitTufWin:TWC2019}
E.~{Leitinger}, F.~{Meyer}, F.~{Hlawatsch}, K.~{Witrisal}, F.~{Tufvesson}, and
  M.~Z. {Win}, ``A belief propagation algorithm for multipath-based {SLAM},''
  \emph{{IEEE} Trans. Wireless Commun.}, vol.~18, no.~12, pp. 5613--5629, Dec.
  2019.

\bibitem{Leitinger2023}
E.~Leitinger, A.~Venus, B.~Teague, and F.~Meyer, ``Data fusion for
  multipath-based {SLAM}: Combining information from multiple propagation
  paths,'' \emph{IEEE Transactions on Signal Processing}, vol.~71, pp.
  4011--4028, 2023.

\bibitem{Venus2024}
A.~Venus, E.~Leitinger, S.~Tertinek, and K.~Witrisal, ``A graph-based algorithm
  for robust sequential localization exploiting multipath for
  obstructed-{LOS}-bias mitigation,'' \emph{IEEE Transactions on Wireless
  Communications}, vol.~23, no.~2, pp. 1068--1084, 2024.

\bibitem{VenusRadar2021}
------, ``A message passing based adaptive {PDA} algorithm for robust
  radio-based localization and tracking,'' in \emph{2021 Proc. IEEE
  RadarConf-21}, 2021, pp. 1--6.

\bibitem{VenLeiTerMeyWit:TWC2024}
A.~{Venus}, E.~{Leitinger}, S.~{Tertinek}, F.~{Meyer}, and K.~{Witrisal},
  ``Graph-based simultaneous localization and bias tracking,'' \emph{{IEEE}
  Trans. Wireless Commun.}, vol.~23, no.~10, pp. 13\,141--13\,158, May 2024.

\bibitem{Koch2008}
J.~W. Koch, ``Bayesian approach to extended object and cluster tracking using
  random matrices,'' \emph{IEEE Transactions on Aerospace and Electronic
  Systems}, vol.~44, no.~3, pp. 1042--1059, 2008.

\bibitem{Granstroem2014}
K.~Granstr{\"o}m, S.~Reuter, D.~Meissner, and A.~Scheel, ``A multiple model
  {PHD} approach to tracking of cars under an assumed rectangular shape,'' in
  \emph{17th International Conference on Information Fusion (FUSION)}, 2014,
  pp. 1--8.

\bibitem{Baum2014}
M.~Baum and U.~D. Hanebeck, ``Extended object tracking with random hypersurface
  models,'' \emph{IEEE Transactions on Aerospace and Electronic Systems},
  vol.~50, no.~1, pp. 149--159, 2014.

\bibitem{BarShalomTCS2009}
Y.~Bar-Shalom, F.~Daum, and J.~Huang, ``The probabilistic data association
  filter,'' \emph{{IEEE} Control Syst. Mag.}, vol.~29, no.~6, pp. 82--100, Dec
  2009.

\bibitem{MeyerTSP2021}
F.~Meyer and J.~L. Williams, ``Scalable detection and tracking of geometric
  extended objects,'' \emph{{IEEE} Trans. Signal Process.}, vol.~69, pp.
  6283--6298, Oct. 2021.

\bibitem{Wielandner2023}
L.~Wielandner, A.~Venus, T.~Wilding, and E.~Leitinger, ``Multipath-based {SLAM}
  for non-ideal reflective surfaces exploiting multiple-measurement data
  association,'' \emph{J. Adv. Inf. Fusion}, vol.~18, pp. 59--77, Dec. 2023.

\bibitem{WieVenWilWitLei:Fusion2024}
L.~Wielandner, A.~Venus, T.~Wilding, K.~Witrisal, and E.~Leitinger, ``{MIMO}
  multipath-based {SLAM} for non-ideal reflective surfaces,'' in \emph{Proc.
  Fusion-2024}, Venice, Italy, Jul. 2024.

\bibitem{Hoher2022}
P.~Hoher, S.~Wirtensohn, T.~Baur, J.~Reuter, F.~Govaers, and W.~Koch,
  ``Extended target tracking with a lidar sensor using random matrices and a
  virtual measurement model,'' \emph{{IEEE} Trans. Signal Process.}, vol.~70,
  pp. 228--239, Dec. 2022.

\bibitem{WildingEUCAP2022}
T.~Wilding, E.~Leitinger, and K.~Witrisal, ``Multipath-based localization and
  tracking considering off-body channel effects,'' in \emph{2022 16th European
  Conference on Antennas and Propagation (EuCAP)}, 2022, pp. 1--5.

\bibitem{WitrisalJWCOML2016}
K.~Witrisal, E.~Leitinger, S.~Hinteregger, and P.~Meissner, ``Bandwidth scaling
  and diversity gain for ranging and positioning in dense multipath channels,''
  \emph{{IEEE} Wireless Commun. Lett.}, vol.~5, no.~4, pp. 396--399, Aug. 2016.

\bibitem{LeitingerJSAC2015}
E.~Leitinger, P.~Meissner, C.~Ruedisser, G.~Dumphart, and K.~Witrisal,
  ``Evaluation of position-related information in multipath components for
  indoor positioning,'' \emph{{IEEE} J. Sel. Areas Commun.}, vol.~33, no.~11,
  pp. 2313--2328, Nov. 2015.

\bibitem{Feldmann2011}
M.~Feldmann, D.~Fr{\"a}nken, and W.~Koch, ``Tracking of extended objects and
  group targets using random matrices,'' \emph{{IEEE} Trans. Signal Process.},
  vol.~59, no.~4, pp. 1409--1420, Sep. 2011.

\bibitem{Wan2000}
E.~Wan and R.~Van Der~Merwe, ``The unscented {K}alman filter for nonlinear
  estimation,'' in \emph{Proc. IEEE ASSPCC 2000}, Aug. 2000, pp. 153--158.

\bibitem{KschischangTIT2001}
F.~Kschischang, B.~Frey, and H.-A. Loeliger, ``Factor graphs and the
  sum-product algorithm,'' \emph{{IEEE} Trans. Inf. Theory}, vol.~47, no.~2,
  pp. 498--519, Feb. 2001.

\bibitem{BarShalom2002EstimationTracking}
Y.~Bar-Shalom, T.~Kirubarajan, and X.-R. Li, \emph{Estimation with Applications
  to Tracking and Navigation}.\hskip 1em plus 0.5em minus 0.4em\relax New York,
  NY, USA: John Wiley \& Sons, Inc., 2002.

\bibitem{Tichavsky1998}
P.~Tichavsky, C.~Muravchik, and A.~Nehorai, ``Posterior {Cramer-Rao} bounds for
  discrete-time nonlinear filtering,'' \emph{{IEEE} Trans. Signal Process.},
  vol.~46, no.~5, pp. 1386--1396, May 1998.

\end{thebibliography}
